\documentclass{webofc}
\usepackage[varg]{txfonts}   
\def\pythia{\textsc{Pythia8}~}
\def\dipsy{\textsc{Dipsy}~}
\usepackage[utf8]{inputenc}

\begin{document}
\title{Rope Hadronization and Strange Particle Production}
%
%

\author{\firstname{Christian} \lastname{Bierlich}\inst{1}\fnsep\thanks{\email{christian.bierlich@thep.lu.se}. Report numbers: LU-TP 17-31, MCnet-17-15. Work done in collaboration with: J. Christiansen, G. Gustafson, L. L\"onnblad and A. Tarasov. Work supported in part by the MCnetITN FP7 Marie Curie Initial Training Network, contract PITN-GA-2012-315877, and the Swedish Research Council (contracts 621-2012-2283, 621-2013-4287 and 2017-00347).}
}

\institute{Department of Astronomy and Theoretical Physics, Lund University\\
	   S\"olvegatan 14A, Lund, Sweden.
          }

\abstract{%
Rope Hadronization is a model extending the Lund string hadronization model to describe environments with many overlapping strings, such as high multiplicity pp collisions or $AA$ collisions. Including effects of Rope Hadronization drastically improves description of strange/non-strange hadron ratios as function of event multiplicity in all systems from $e^+ e^-$ to $AA$. Implementation of Rope Hadronization in the MC event generators \dipsy and \pythia is discussed, as well as future prospects for jet studies and studies of small systems. 	
}
\maketitle
\section{Introduction}

The Lund string model for hadronization \cite{Andersson:1983jt,Andersson:1979ij} is based on the "string" picture of QCD, where the confined colour field between a $q\bar{q}$ pair moving apart is modelled as a classical string, with tension $\kappa \approx 1 $GeV/fm. When it becomes energetically favorable for the string to break into smaller pieces, it breaks according to the Schwinger equation \cite{Schwinger:1951nm} (modified to take transverse degrees of freedom into account) with the probability:
\begin{equation}
	\label{eq:sch}
	\frac{d\mathcal{P}}{dp_\perp} \propto \kappa \exp\left( -\frac{\pi m^2_\perp}{\kappa} \right).
\end{equation}
Directly from eq. (\ref{eq:sch}), one obtains suppression of hadrons with strange quark content wrt. hadrons without strange quarks. Factorizing the $p_\perp$ dependence out, the suppression of strange quarks in the string breaking becomes:
\begin{equation}
	\label{eq:sup}
	\rho = \frac{\mathcal{P}_s}{\mathcal{P}_{u,d}} = \exp\left(-\frac{\pi(m^2_s - m^2_u)}{\kappa}\right).
\end{equation}
Taking eq. (\ref{eq:sup}) at face value, inserting constituent quark masses of $m_u \approx 0.3$ GeV and $m_s \approx 0.5$ GeV, 
one obtains a suppression factor of $\rho \approx 0.08$, which is too small to describe data. In practise one therefore treats $\rho$ as a free parameter, fitted to data from $e^+ e^-$ collisions \cite{Hamacher:1995df}, along with the rest of the parameters of the fragmentation model. Such sets of parameters are usually referred to as a "tune", taken as a hard constraint on parameter values when making predictions for other collision systems such a pp, p$A$ or $AA$. One example is the Monash tune \cite{Skands:2014pea} for \pythia \cite{Sjostrand:2014zea}, from which it can be seen that the string model does a good job of described hadron flavours in an $e^+ e^-$ environment.

\begin{figure}[h]
\centering
\includegraphics[width=0.5\textwidth,clip]{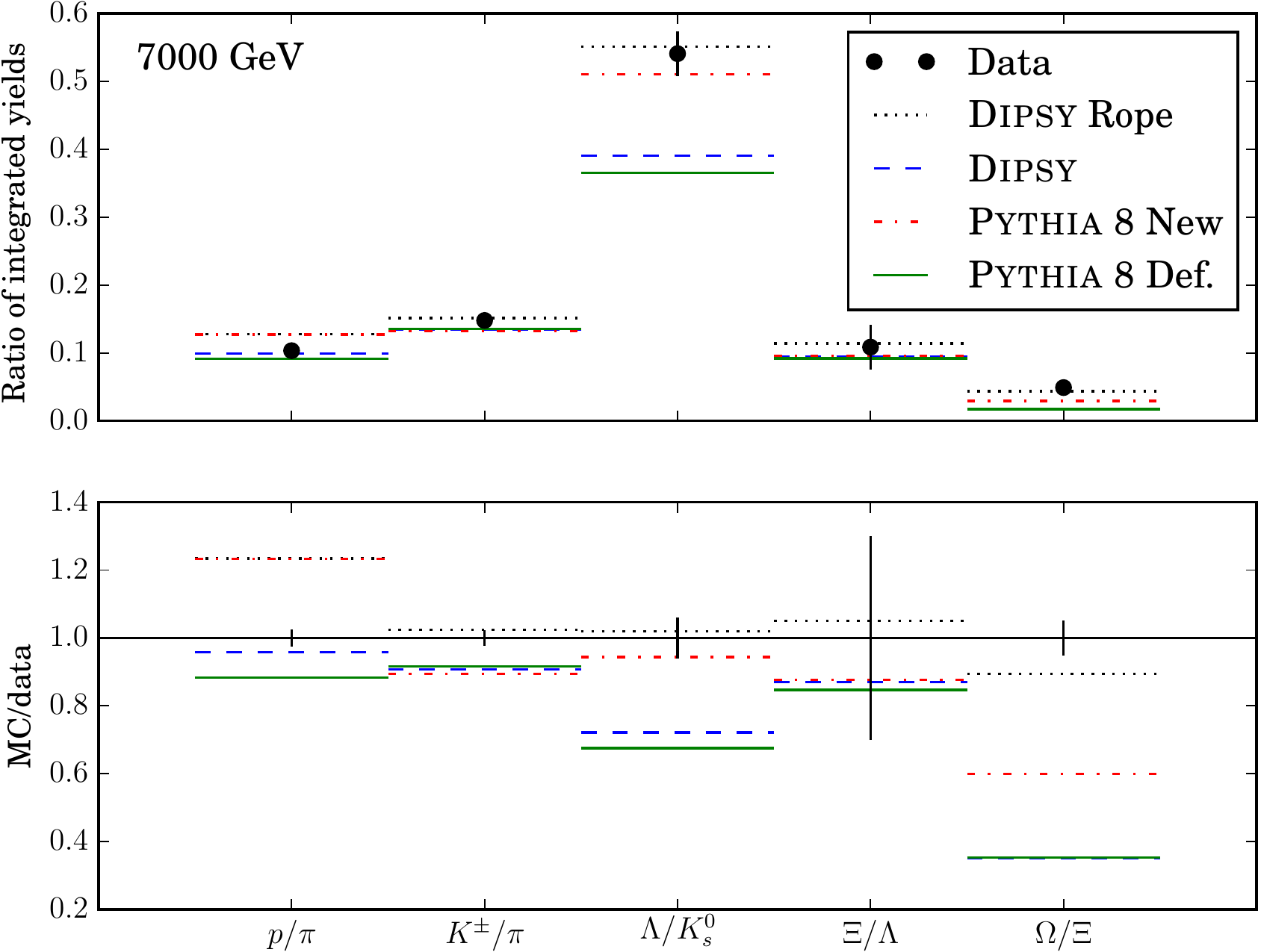}
	\caption{Default string hadronization and rope hadronization compared to inclusive ratios of identified particles in pp collisions at $\sqrt{s} = 7$ TeV.}
\label{fig:ppratios}       
\end{figure}

Turning to pp collisions, the picture is somewhat changed. In figure \ref{fig:ppratios} (taken from ref. \cite{Bierlich:2015rha}), it is shown how the unmodified string picture does not hold even for inclusive ratios of identified particles, when emphasis is put on strangeness. The theoretical curves "\dipsy" and "\pythia def." represent unmodified string hadronization, "\pythia new" represents a modified colour reconnection scheme in \pythia described in ref. \cite{Christiansen:2015yqa} (which shall not be mentioned further here), while "\dipsy rope" represents the Rope Hadronization model, which is the subject of these proceedings.

\section{Rope Hadronization}
One crucial assumption behind string hadronization, as described in the previous section, is that strings are allowed to fragment independently of each other. It was early noted that in the dense environment of nuclear collisions, strings might act together coherently to form a stronger "rope" \cite{Biro:1984cf,Bialas:1984ye}, which would then hadronize with a larger, effective string tension. This idea has been followed by several authors \cite{Kerman:1985tj,Gyulassy:1986jq,Andersson:1991er,Braun:1991dg,Braun:1993xw,Amelin:1994mc,Armesto:1994yg,Sorge:1992ej,Bleicher:2000us,Soff:2002bn}, also for $pp$ collisions. In ref. \cite{Bierlich:2014xba}, the rope hadronization model was implemented in the parton shower event generator \dipsy \cite{Flensburg:2011kj} (using \pythia for hadronization), and in ref. \cite{Bierlich:2016faw} as a direct plug-in to the \pythia MC. 
The key point of the Rope Hadronization model, is the increase of local string tension. In the \dipsy and \pythia implementations this is calculated by calculating the transverse-space overlap of string pieces. Denoting the number of string pieces parallel with the breaking string $p$, and anti-parallel string pieces $q$, one finds that the effective string tension ($\tilde{\kappa}$) for a break-up becomes:
\begin{equation}
 \label{eq:kappaeff}
	\tilde{\kappa} = \frac{2p+q+2}{4} \kappa.
\end{equation}
This result is confirmed by lattice calculations indicating that the string tension of an SU(3) multiplet structure scales with the secondary Casimir operator of the multiplet \cite{Bali:2000un}. A key difference between this Rope Hadronization model and other earlier models, is that the effective string tension in eq. (\ref{eq:kappaeff}) is calculated as the difference of the available energy in the full multiplet and the multiplet reached by breaking one string in the rope (\textit{i.e} replacing $p \rightarrow p - 1$). If one instead uses the full multiplet to calculate the effective string tension, effects becomes much too large.
Replacing $\kappa$ with $\tilde{\kappa}$ in eq. (\ref{eq:sup}), the $s$-suppression becomes $\tilde{\rho} = \rho^{\kappa/\tilde{\kappa}}$, approaching unity as $\tilde{\kappa}$ grow large, when many strings overlap (in a pp collision at LHC, $\tilde{\rho}$ reaches values of up to $\approx 3\rho$).

One should note in particular the importance of multi-strange particles. Especially the $\phi$-meson, consisting of two $s$-quarks, is important, as its mass is quite similar to the mass of the proton, making it an excellent laboratory for studying whether enhancement with increasing multiplicity is really due to strangeness enhancement, or if it is due to a mass effect, as indicated by some thermodynamical calculations \cite{Vislavicius:2016rwi}.

\section{Results}
\begin{figure}[h]
\centering
\includegraphics[width=0.7\textwidth,clip]{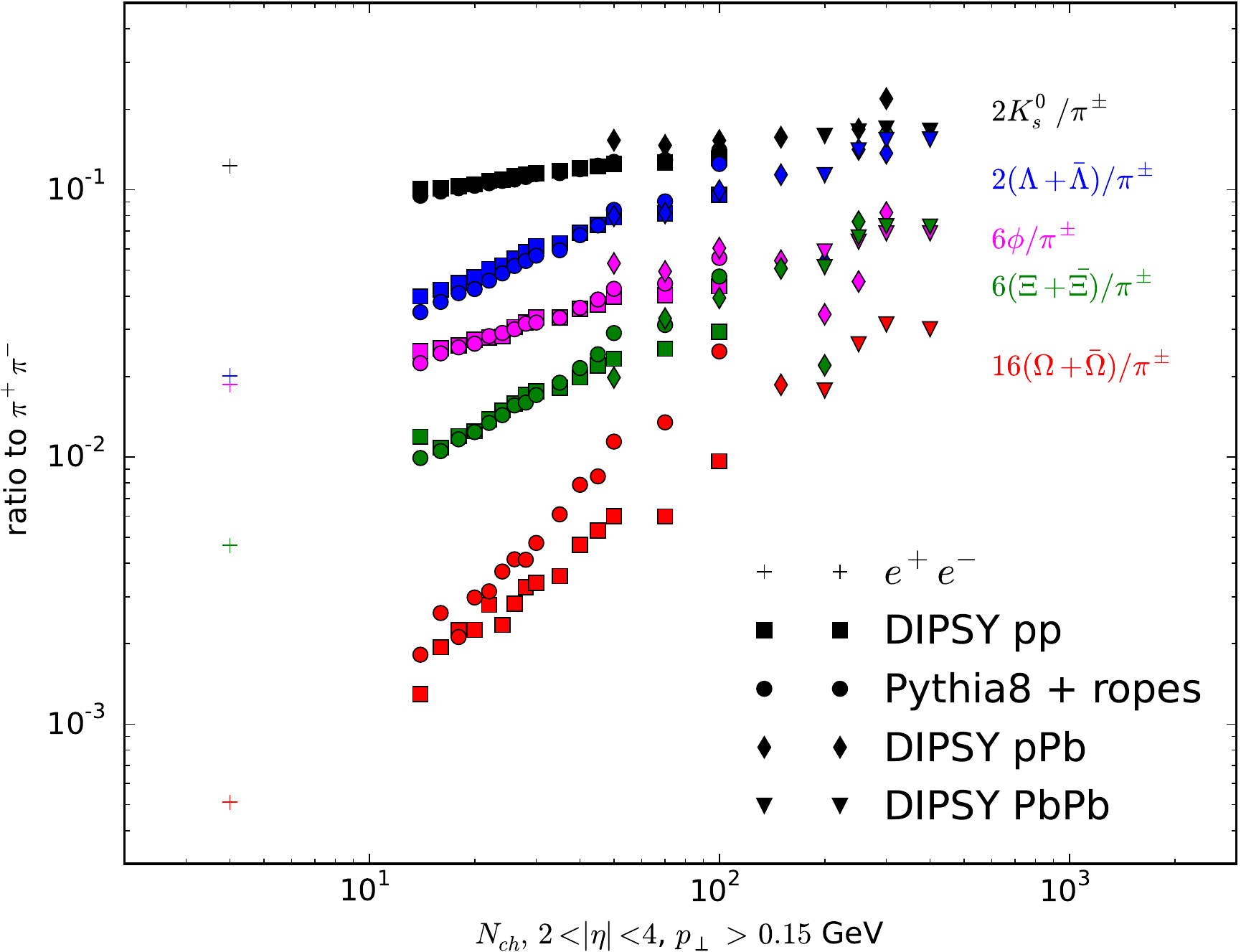}
	\caption{Ratios of strange particles to pions in $e^+ e^-$, pp, pPb and PbPb collisions as function of multiplicity (centrality) with Rope Hadronization as implemented in the \pythia and \dipsy MC event generators.}
\label{fig:sresults}       
\end{figure}

In figure \ref{fig:sresults}, predicted enhancement for several strange and multi-strange hadrons as function of event multiplicity is shown. Results are shown both for the original \dipsy implementation, as well as the \pythia implementation (the latter for pp only). The figure can be compared to experimental results obtained by ALICE \cite{ALICE:2017jyt}. Rope Hadronization can be seen to describe strangeness in all systems from $e^+ e^-$ to PbPb well.
As noted above, the $\phi/\pi$ ratio is expected to rise, even in pp. As this can serve as a potential discriminator between Rope Hadronization and thermodynamical calculations, the measurement is theoretically well motivated.

\section{Outlook}
Strangeness enhancement has historically been seen as an important observable for QGP formation. In these proceedings it has been shown that the Rope Hadronization mechanism -- a microscopic model assuming neither a deconfined nor thermalized plasma -- can describe strangeness in collision systems from $e^+ e^-$ to $AA$. The question which begs to be answered is if such a microscopic model can describe other features usually ascribed to plasma formation, such as flow and jet quenching. Regarding flow, ongoing efforts \cite{Bierlich:2016vgw} aims to investigate whether the transverse pressure gradient which is naturally built up due to string overlaps, can be translated into flow. The effects of Rope Hadronization on jets in pp collisions is also worth investigating further \cite{Mangano:2017plv}, though the Rope Hadronization model is develeoped primarily with soft effects in mind. Finally it is worth mentioning the prospects of looking further into effects even smaller systems, such as central diffractive pp collisions or $e^+ e^-$ collisions at an FCC-ee machine \cite{Bierlich:2017ulc}. The absence of a coloured initial state in the latter makes it a valuable precision tool for determining whether collective effects arise from initial state correlations or final state interactions.

%
\bibliography{refs}
%
%
%
%

\end{document}